\documentclass[sigconf]{acmart} 

\renewcommand\footnotetextcopyrightpermission[1]{} 
\usepackage{color} 
\usepackage{tikz}
\usepackage{subfig}
\usepackage{comment}
\usepackage{booktabs} 
\usepackage{makecell}
\usepackage{multirow}
\usepackage{graphicx} 
\usepackage{comment} 
\usepackage{balance}

\acmDOI{10.475/123_4}
\acmISBN{123-4567-24-567/08/06}
\acmConference[WebSci'18]{10th ACM Conference on Web Science}{May 2018}{Amsterdam, the Netherlands} 
\acmYear{2018}
\copyrightyear{2018}
\acmArticle{4}
\acmPrice{15.00}

\newcommand{\q}[1]{\lq\lq{}{}#1\rq\rq{}{}}

\begin{document}

\copyrightyear{2018} 
\acmYear{2018} 
\setcopyright{acmcopyright}
\acmConference[WebSci'18]{10th ACM Conference on Web Science}{May 27--30, 2018}{Amsterdam, Netherlands}
\acmBooktitle{WebSci'18: 10th ACM Conference on Web Science, May 27--30, 2018, Amsterdam, Netherlands}
\acmPrice{15.00}
\acmDOI{10.1145/3201064.3201076}
\acmISBN{978-1-4503-5563-6/18/05}

\title{Viewpoint Discovery and Understanding in Social Networks}

\author{Mainul Quraishi}
\affiliation{
  \institution{L3S Research Center,\\Leibniz University of Hannover}
  \city{Hannover, Germany}
}
\email{quraishimainul@gmail.com}

\author{Pavlos Fafalios}
\orcid{0000-0003-2788-526X}
\affiliation{
  \institution{L3S Research Center,\\Leibniz University of Hannover}
  \city{Hannover, Germany}}
\email{fafalios@L3S.de}

\author{Eelco Herder}
\affiliation{
  \institution{Radboud University}
  \city{Nijmegen, the Netherlands}}
\email{eelcoherder@acm.org}

\begin{abstract}
 The Web has evolved to a dominant platform where everyone has the opportunity to express their opinions, to interact with other users, and to debate on emerging events happening around the world. On the one hand, this has enabled the presence of different viewpoints and opinions about a $-$ usually controversial $-$ topic (like Brexit), but at the same time, it has led to phenomena like media bias, echo chambers and filter bubbles, where users are exposed to only one point of view on the same topic. Therefore, there is the need for methods that are able to {\em detect} and {\em explain} the different viewpoints.
 In this paper, we propose a graph partitioning method that exploits social interactions to enable the discovery of different communities (representing different viewpoints) discussing about a controversial topic in a social network like Twitter. To explain the discovered viewpoints, we describe a method, called {\em Iterative Rank Difference} (IRD), which allows detecting descriptive terms that characterize the different viewpoints as well as understanding how a specific term is related to a viewpoint (by detecting other related descriptive terms). The results of an experimental evaluation showed that our approach outperforms state-of-the-art methods on viewpoint discovery, while a qualitative analysis of the proposed IRD method on three different controversial topics showed that IRD provides comprehensive and deep representations of the different viewpoints. 
\end{abstract}

\ccsdesc[500]{Information systems~Clustering and classification}

\keywords{Viewpoint discovery; Viewpoint understanding; Social networks}

\maketitle

\section{Introduction}

Social Media has now emerged as the dominant platform to discuss and comment on breaking news and noteworthy events that are happening around the world. 
On Twitter, for example, every second around 6,000 tweets are posted, which corresponds to over 350,000 tweets per minute, 500 million tweets per day and around 200 billion tweets per year\footnote{\url{http://www.internetlivestats.com/twitter-statistics/} (April 4, 2018)}.

Although such networking services facilitate the expression of diverse opinions and viewpoints about particular topics, it has led to the widespread observation of phenomena such as \textit{media bias} and the so-called \textit{echo chamber} effect, where users are increasingly exposed to conforming opinions.

Different viewpoints and bias have been observed in tweets from regular newspaper publishers, and previous research has shown that based on these tweets it is possible to position these publishers with respect to their stance on economic and personal issues \cite{Elejalde:2017:NRP:3078714.3078724}. Similarly, it has been shown that there is bias in the selection of which parts of ongoing events are reported and in which depth they are covered \cite{tran2015detecting}, arguably due to differences in national or regional interests, differences in political or societal orientation, and other factors.

It has been also shown that individuals generally read publications that are ideologically quite similar and are usually exposed to only one side of the political spectrum \cite{flaxman2016filter}. Due to these self-selected echo chambers and the observation that users with similar interests and opinions are more likely to be friends on social media (the so-called \textit{homophily effect}) \cite{Aiello:2012:FPH:2180861.2180866}, polarized discussions of user groups with sometimes very different viewpoints are very common on social media.

As a large part of the population is nowadays exposed to biased, filtered or fake news, and actively participates in discussions with like-minded, there are real social costs and effects, as observed in the 2016 US election \cite{allcott2017social}. In order to understand ongoing political or societal debates, or to analyze and interpret the course of historical events retrospectively, there is the need for methods that are able to detect the different viewpoints pertaining a topic and to understand what they are about. 

In a nutshell, this paper makes the following contributions:
\begin{itemize}
    \item We propose the use of a popular graph partitioning method that exploits social interactions to cluster the users of a social network based on their viewpoint on a controversial topic. Contrary to existing works that follow a similar approach, 
    the proposed method can be applied in cases with unknown number of viewpoints and also enables the detection of noisy groups of users that do not represent clear viewpoints. 

    \item To obtain a deeper understanding of a viewpoint, we introduce a method, called {\em Iterative Rank Difference (IRD)}, which is based on the iterative use of an automatic term recognition algorithm. The proposed method can automatically identify descriptive terms that characterize the different viewpoints and also allows understanding how one or more specific terms are related to a viewpoint. 
    
    \item We present evaluation results which showcase that: i) the proposed viewpoint discovery method outperforms existing state-of-the-art topic models, ii) IRD provides comprehensive descriptions of the different viewpoints as well as of the terms that characterize them.
\end{itemize}

The remainder of this paper is organized as follows: 
Section \ref{sec:rw} reports related works and the difference of our approach. 
Section \ref{sec:modeling} models the problems of {\em viewpoint discovery} and {\em viewpoint understanding}.
Section \ref{sec:discovery} details the proposed {\em viewpoint discovery} method. 
Section \ref{sec:understanding} describes the proposed {\em viewpoint understanding} method. 
Section \ref{sec:evaluation} presents evaluation results. 
Finally, Section \ref{sec:conclusion} concludes the paper and discusses interesting directions for future research.

\section{Related Work}
\label{sec:rw}

Given a controversial topic, for example an issue like {\em climate change} or an entity like {\em Donald Trump}, viewpoint discovery aims to find the different viewpoints expressed about the topic in a set of documents or in a social network.
This task can be considered a sub-task of Opinion Mining, which aims to analyze opinionated documents and to infer properties such as subjectivity or polarity\footnote{The survey in \cite{pang2008opinion} provides a general review on Opinion Mining and Sentiment Analysis.}. Discovering the different viewpoints usually involves (or is performed in parallel with) two other tasks: i) grouping the corresponding documents or users based on their viewpoints; ii) finding descriptive terms that characterize the viewpoints. 

Viewpoint discovery has been applied to both documents and social media data. Below, we report works for both types of data sources.

\subsection{Viewpoint Discovery in Documents}

Paul and Girju \cite{paul2010two} introduced a model, called Topic-Aspect Model (TAM), that jointly considers topics and aspects (where aspects can be interpreted as viewpoints in our case). TAM decomposes each document into a mix of topics that are characterized by a multinomial distribution over words and also exploits a second mix of aspects that affect the topics. To describe the viewpoints pertaining to a topic, \cite{paul2010summarizing} proposed a contrastive viewpoint summarization framework which is based on TAM and is able to find phrases that best reflect the different viewpoints.

Thonet et al \cite{thonet2016vodum} proposed the Viewpoint and Opinion Discovery Unification Model (VODUM), an unsupervised topic model based on LDA \cite{blei2003latent} that leverages parts of speech to jointly discover viewpoints, topics, and opinions in a text (helping to discriminate between topic and opinion words). 

Al Khatib et al \cite{al2012automatic} studied the slightly different problem on how to automatically detect the differences in viewpoints between two articles in Wikipedia, where one article describes the subject matter in a more positive or negative way than the other. A statistical classifier is trained to predict the viewpoint score of a document, which reflects how positive or negative the viewpoint is. 

Another interesting work studied the prediction and analysis of the ideological stance of legislators \cite{nguyen2015tea}. The proposed topic model integrates regression techniques to estimate real-valued ideal points and can also extract and provide a vocabulary that characterizes the ideological discourse.

\subsection{Viewpoint Discovery in Social Media}

Given a controversial topic as well as the texts and interactions from the users discussing it on a social network, Thonet et al \cite{Thonet2017Users} propose an unsupervised topic model, called Social Network Viewpoint Discovery Model (SNVDM), which jointly identifies subtopics, viewpoints and the discourse pertaining to the different subtopics and viewpoints. To infer the viewpoints, SNVDM exploits the {\em homophily} phenomenon (the tendency of individuals to associate and bond with similar others) and relies on the social interactions among the users. Moreover, to account for the sparsity of social networks and the weak connections between users, the authors extend SNVDM into the SNVDM-GPU model which leverages the Generalized P\'olya Urn (GPU) scheme \cite{mahmoud2008polya}.

Under the same assumption that social networks are homophilic, Barber\'{a}
\cite{barbera2014birds} proposes a Bayesian Spatial Following model
that groups Twitter users along a common ideological dimension based on who they follow. 

Ren et al \cite{ren2016time} describe a dynamic modeling strategy to infer contrastive opinions in a given stream of multilingual social texts by jointly modeling topics, entities and sentiment labels. However, contrast is based only on sentiment (positive, negative, neutral), not on viewpoints, and thus this approach actually models and infers sentiment polarity and partisanship.

Joshi et al \cite{joshi2016political} present a topic model, called Political Issue Extraction (PIE), that estimates word-specific distributions denoting political issues and positions from an unlabeled dataset of tweets. The model uses affiliation information of political users as well as Twitter timelines of both political and non-political users. In the same context, 
\cite{conover2011political} investigated how social media shape the networked public sphere and facilitate communication between communities with different political orientations. The authors showed that mentions and retweets, the two major mechanisms for public political interaction on Twitter, induce distinct network topologies: the retweet network is highly polarized, while the mention network is not.

Sachan et al \cite{sachan2014spatial} study the different yet related problem of detecting communities on Twitter. The proposed topic model, called Social Network Latent Dirichlet Allocation (SN-LDA), combines both text and social interactions. However, such methods focus on discovering topical communities (for instance, users discussing about football) and not different viewpoints expressed on a common controversial topic. 
A similar community detection technique was combined with topic modeling to  characterize opinions about human papillomavirus vaccines (HPV) on Twitter \cite{surian2016characterizing}. Papadopoulos et al. \cite{papadopoulos2012community} provide a comprehensive literature review of community detection approaches applied in social media. 

There are also works that have applied traditional machine learning (supervised) classifiers to automatically identify the political affiliation and viewpoint of social media users, such as Naive Bayes \cite{fang2015topic}, Support Vector Machine \cite{cohen2013classifying}, Decision Trees \cite{pennacchiotti2011democrats} or Neural Networks \cite{rao2016actionable}. However, although such models usually achieve high precision, they need training examples which may be very difficult to obtain. 

Another related line of research has applied topic models for viewpoint discovery in forums \cite{qiu2013latent,qiu2013modeling}. These works exploit the threaded nature of forum posts which though is not the case in social networks like Twitter where threaded interactions are scarce. 

Finally, there is a number of works that tackle the related problem of {\em controversy detection}. This task focuses on distinguishing whether a topic of online discussion is controversial or not. Works in this field handle the problem from different perspectives, e.g., by analyzing news articles  \cite{choi2010identifying,mejova2014controversy}, by exploiting the social media \cite{popescu2010detecting,garimella2016quantifying}, or by focusing on Wikipedia pages \cite{borra2015societal,dori2015automated}.
As regards social media, controversy detection has been studied on Twitter using both supervised \cite{popescu2010detecting} and unsupervised \cite{garimella2016quantifying} methods.
Popescu and Pennacchiotti \cite{popescu2010detecting} proposed and evaluated three regression machine learning models for identifying controversial events in Twitter.
The models make use of a variety of features, including linguistic, structural, sentiment and news-based ones. 
On the other hand, Garimella et al. \cite{garimella2016quantifying} developed a framework to identify controversy without prior domain-specific knowledge about the topics in question. 
The framework builds a conversation graph about a topic, partitions the graph to identify potential sides of the controversy, and measures the amount of controversy by analyzing characteristics of the graph.

\subsection{Our approach}
Compared to the existing works, in this paper we also exploit social interactions (like \cite{Thonet2017Users} and \cite{conover2011political}) and build an interaction graph where nodes correspond to users and edges to endorsements among the users. To detect the different viewpoints, actually the groups of users that represent the viewpoints, we make use of a popular multi-level graph partitioning algorithm, inspired by the work in \cite{garimella2016quantifying} which applies the same algorithm for the problem of controversy detection and quantification. However, contrary to this work, we propose the use of the quality metric {\em conductance} for deciding on the clusters that hold different viewpoints. 
In this way, our method can be also applied in cases of unknown number of viewpoints as well as for the identification of noisy groups of users that do not represent a clear viewpoint. 

With respect to the problem of {\em viewpoint understanding}, in order to provide a deep understanding of a viewpoint, we propose the iterative use of an automatic term recognition algorithm that is based on \textit{rank difference} \cite{kit2008measuring}. To our knowledge, this kind of algorithms has not been previously applied to a similar problem.

\section{Problem Modeling}
\label{sec:modeling}

In our context, a {\em controversial topic} is a subject of public interest discussed in a social networking service where its users can post texts and also interact with each other. A controversial topic can be a specific event (e.g., {\em US Election 2016}), a general issue (e.g., {\em abortion} or {\em gun control}), or even an entity (e.g., {\em Donald Trump} or {\em Palestine}). 

For a controversial topic $t$, let $P$ be the set of all texts (posts) related to $t$, posted in a social networking service by a set of users $U$, in a specific time period. We can gather these texts by querying the social network using specific keywords that describe the topic. Let also $W$ be the vocabulary of $P$, i.e., all terms/words that exist in the texts of $P$.
Now, let $I$ be the set of {\em endorsement} interactions among the users (e.g., likes or retweets) where each such interaction is associated with a post in $P$ and two users in $U$ (i.e., the interactions are related to the topic). For a post $p \in P$ and two users $u_1, u_2 \in U$, we can represent an interaction $I_i \in I$ as a triple of the form $\langle u_1, p, u_2 \rangle$, where the user $u_1$ is the initiator (or sender) of the interaction, user $u_2$ is the receiver, and $p$ is the mean (or channel) of interaction (e.g., $u_1$ likes the text $p$ posted by $u_2$). 

We now model the problems of {\em viewpoint discovery} and {\em viewpoint understanding} as follows: 
\begin{itemize}
    \item {\em Viewpoint discovery:} compute a set of $k$ user groups $G = \{G_1, \dots, G_k\}$, where each group $G_i \subset U$ corresponds to a different viewpoint.
    
    \item {\em Viewpoint understanding:} for each group $G_i \in G$, 
    compute a set of top-$n$ descriptive terms $W_i \subset W$ that characterize the viewpoint.
    In addition, for a descriptive term $w \in W_i$, compute a new set of descriptive terms $W'_{i} \subset W$ that characterize $w$ in the context of the viewpoint $G_i$. 
\end{itemize}

\section{Viewpoint Discovery}
\label{sec:discovery}

Similar to \cite{garimella2016quantifying} and \cite{conover2011political}, we  build an {\em interaction graph} $\mathcal{G}(U, I)$, where nodes correspond to users and edges correspond to endorsements among the users. In our case, an edge connecting two users is undirected and exists if there is a sufficient number of interactions that involve them. In our case studies and experiments we draw an edge when there are at least 2 endorsements between the two users (similar to \cite{garimella2016quantifying}). This number seems to be sufficient for indicating a common viewpoint on a topic, while it also reduces the noise in the considered data.

Below, we first discuss how we decided on the clustering method to use in our problem (Section \ref{subsec:selectingMethod}), and then describe the method we propose for detecting the clusters that hold different viewpoints (Section \ref{subsec:detectingViewpoints}).

\subsection{Choosing the proper clustering method}
\label{subsec:selectingMethod}
We need a clustering method which can ensure sparse connections among the different clusters. The intuition for this requirement is the following: the absence of an edge between two users does not imply different viewpoints, however users with different viewpoints will probably not endorse each other. Thus, contrary to classic community detection algorithms, ensuring dense connections within the same cluster is not a requirement for our problem. 
In addition, since the network can be quite large, containing thousands or even millions of users, we should avoid approaches that require pairwise node comparisons.

We studied the applicability of different network clustering algorithms: 
\begin{itemize}
    \item Node similarity
    \item The Louvain method (community detection)
    \item Multi-Level Graph Partitioning (MLGP)  
\end{itemize}

Node similarity algorithms aim to find nodes that are topologically similar, i.e., they share many of the same neighbors.  
Various similarity measures can be used, like cosine similarity, Pearson coefficient, or the Euclidean distance \cite{newman2010networks}. 
However, similarity is measured between each pair of nodes, which makes their computation very time consuming for networks with large number of nodes.

The goal of community detection algorithms is the detection of the dense portions of the network. There is a plethora of algorithms for this kind of problem, like hierarchical clustering and modularity maximization. We tried one of the most popular modularity maximization algorithms called the Louvaine Method \cite{blondel2008fast}. 
Modularity is a measure designed to measure the strength of division of a network into modules (groups or communities), where networks with high modularity have two main characteristics: i) dense connections between the nodes within modules, and ii) sparse connections between nodes in different modules \cite{newman2004analysis}. This means that, if there is no edge connecting two nodes in the same cluster, modularity get decreases as a penalty, for ensuring that the detected clusters are very dense. 
However, as we have already discussed, sparsity is common in large social networks: no edges between two users does not imply different viewpoints.

The goal of Multi-Level Graph Partitioning (MLGP) algorithms is the detection of partitions with a minimum \q{cut}, i.e., a minimum number of edges connecting different clusters. 
In such a method, the network is divided into a predetermined number of parts, 
chosen such that the cut is minimized. 
This approach is suitable in our case because it ensures sparse connections between nodes in different clusters.
To avoid getting partitions of very different sizes (e.g., one partition containing only one node, and one containing all the other nodes),
the size of the partitions need to be also taken into consideration. 
This can be done by computing the degree of the nodes that are inside each cluster. 
This normalized measure can be then used as the optimization criterion for finding the best solution. 
To find the optimal solution, we exploited a {\em multi-level graph partitioning} algorithm (of the METIS software package \cite{karypis1995metis}) which falls under the type of heuristic algorithms.

\subsection{Detecting the viewpoints}
\label{subsec:detectingViewpoints}

\subsubsection{Choosing the right quality metric}
We now need a quality metric for deciding on the clusters that hold different viewpoints. We experimented with the following measures \cite{almeida2011there}:
\begin{itemize}
    \item Modularity
    \item Silhouette index
    \item Coverage
    \item Performance
    \item Conductance
\end{itemize}
As we have already explained (cf. Section \ref{subsec:selectingMethod}), {\em modularity} should be avoided for sparse graphs, since it gives high scores to clusters with sparse internal connections.
The {\em Silhouette index} considers the similarity and dissimilarity between the nodes which in our case can be measured by their shortest path. However, this requires finding the shortest path for all pairs of nodes, which is impractical for very large sparse graphs. 
The {\em coverage} of a clustering is given as the fraction of the weight of all intra-cluster edges with respect to the total weight of all edges in the whole graph. Higher values of coverage means that there are more edges inside the clusters than edges linking different clusters, which translates to a better clustering. However, this measure does not consider the size of a cluster, which makes it unsuitable for our case, since it can give high scores to a clustering having a large number of very small clusters. 
{\em Performance} counts the number of internal edges in a cluster along with the edges
that don't exist between the cluster's nodes and other nodes in the graph. Higher values indicate that a cluster is both internally dense and externally sparse. 
Similar to {\em modularity}, this measure is not suitable for large sparse graphs \cite{almeida2011there}.

Finally, the {\em conductance} of a cut is the ratio between the size of the cut and the minimum cluster volume \cite{kannan2004clusterings}. The volume of a cluster is the total number (or the sum of the weights) of the edges with at least one endpoint in the cluster. 
Specifically, the conductance of a cluster of users $G_i \in G$ can be computed as follows: 
\begin{equation}
\label{condForm}
   conductance(G_i) = \frac{ \sum_{u \in G_i}{\sum_{v \notin G_i}{weight(u,v)}} } {minimum(volume(G_i),volume(\overline{G_i}))}
\end{equation}
where $\overline{G_i} = U \setminus G_i$ and:
\begin{equation}
volume(G_i) = \sum_{u \in G_i}{\sum_{v \in U}{weight(u,v)}}
\end{equation}

In simple words, the formula computes the percentage of edges starting from nodes within the cluster that point to nodes outside the cluster. Since conductance considers both the cut size and the volume of the clusters, this measure is suitable in our problem: a large cluster having a small cut (small number of edges pointing to nodes outside the cluster) will get a low conductance score, making it a good candidate for representing a concrete viewpoint.

\subsubsection{Deciding on the clusters that hold different viewpoints}
To find the clusters that hold different viewpoints, we inspect the conductance of the clusters for different graph partitioning approaches (values of $k$). If a cluster has a large conductance value, this means that the cluster has a strong connection with one or more of the other clusters and thus does not hold a different and clear viewpoint. In general, large conductance values is an indication of poor clustering. Hence, an ideal partitioning provides clusters with very low conductance values. In this case, the network exhibits a highly segregated structure, with limited connectivity among the different clusters, meaning that each cluster represents a different viewpoint. 

\begin{figure}[h]
    \includegraphics[scale=.47]{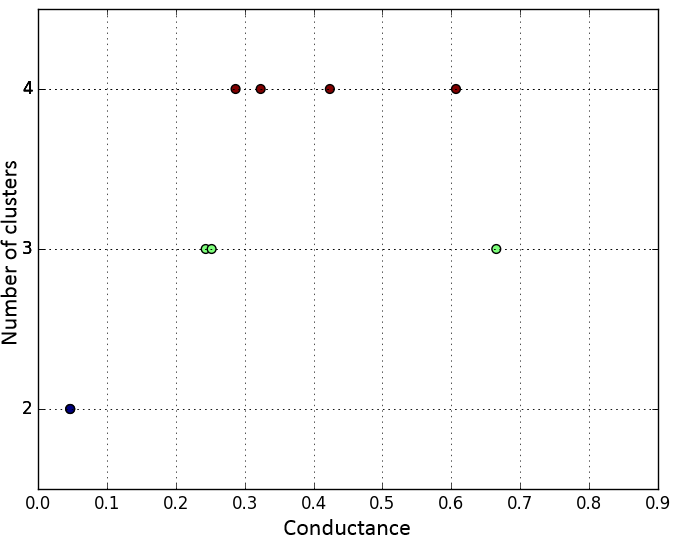}
    \caption{Conductance values for different number of clusters (value of $k$), for the topic {\em \q{2014 Scottish Independence Referendum}}.}
    \label{fig:CondInd}
\end{figure}

Figure \ref{fig:CondInd} depicts the conductance values for different graph partitioning scenarios for the controversial topic {\em 2014 Scottish Independence Referendum}, using the dataset provided in \cite{brigadir2015analyzing} which contains only supporters of \q{yes} and \q{no} (more in Section \ref{sec:evaluation}).
We notice that for $k=2$, the two clusters have very low conductance value. Notice that, since the considered graph is undirected, the clusters have always the same conductance value for $k=2$ (same number of edges connecting the two clusters, same minimum volume of the two clusters). For $k=3$, there are two clusters with much bigger (compared to $k=2$) conductance value, while the third cluster has a very large value meaning that this cluster has a strong connection with one or both of the other two clusters.
Thus, we should partition the graph using $k=2$ and select both clusters as representing different viewpoints.

\begin{figure}[h]
    \includegraphics[scale=.47]{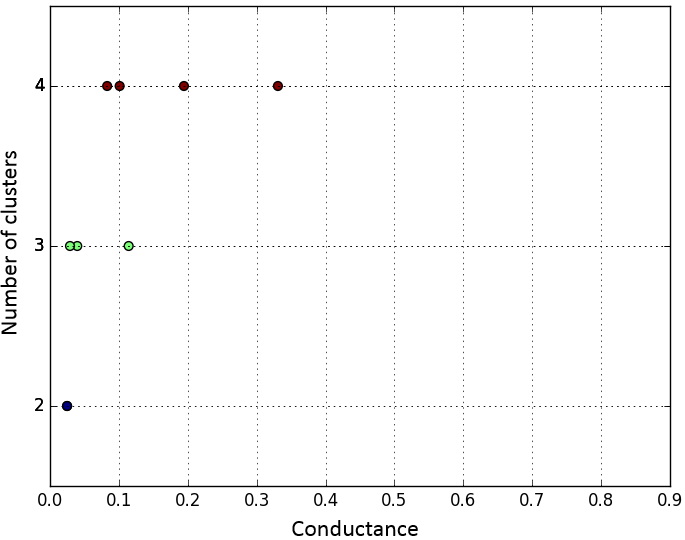}
    \caption{Conductance values for different number of clusters (value of $k$), for the topic {\em \q{2016 US Presidential Election}}.}
    \label{fig:CondUSEl}
\end{figure}

\begin{figure}[h]
    \subfloat[$k=2$]{
        \includegraphics[scale=.38]{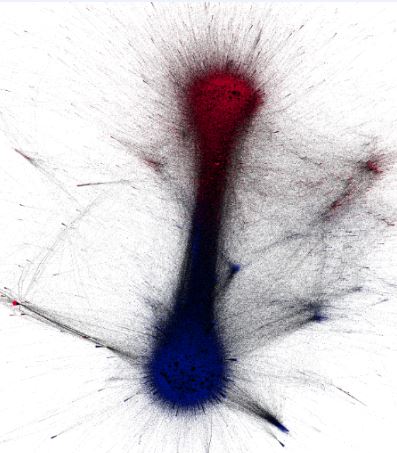}
        \label{fig:us_election_force_atlas_2}}
    \subfloat[$k=3$]{
        \includegraphics[scale=.38]{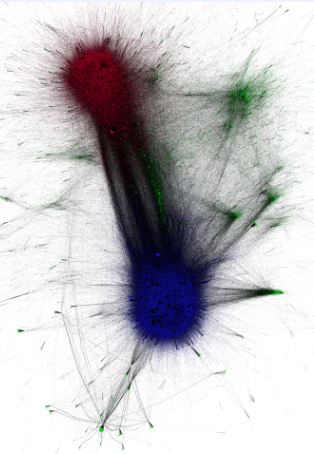}
        \label{fig:us_election_force_atlas_3}}
    \caption{Force-directed network visualization of the topic {\em \q{2016 US Presidential Election}} for different number of clusters (value of $k$). Colors correspond to different viewpoints detected by the proposed MLGP method.}
    \label{fig:usElectionClusters}
\end{figure}

The conductance plot can also indicate cases where a bigger value of $k$ should be selected even if not all $k$ clusters represent viewpoints, because then we can filter out noisy data (i.e., users with no clear viewpoint). This can result in a better selection of the groups of users that hold different viewpoints. 
For instance, Figure \ref{fig:CondUSEl} depicts the conductance plot for the topic {\em 2016 US Presidential Election} (using a manually created dataset, more in Section \ref{sec:evaluation}). Here we should better decide to partition the graph using $k=3$  because two of the clusters have very low conductance values (similar to the case of $k=2$) while the third one has a much larger value. Thus, in this case we can consider as viewpoints only the two clusters with the small conductance values and ignore the third cluster.
To validate this, Figure \ref{fig:usElectionClusters} shows the network clustering for $k=2$ (left) and $k=3$ (right) using a force-directed layout algorithm \cite{jacomy2014forceatlas2}, where colors correspond to different viewpoints detected by our method. 
We notice that, for $k=2$ there are two large well-concentrated clusters that correspond to two different viewpoints, but also several other very small clusters assigned to these two viewpoints. These small clusters is an indication of \q{noisy} groups of users with no clear viewpoint. 
For $k=3$ we notice that there are again two large well-concentrated clusters corresponding to two different viewpoints, while the third viewpoint (in green color) is very scattered and consists of several very small clusters. 

\begin{figure}[h]
    \includegraphics[scale=.50]{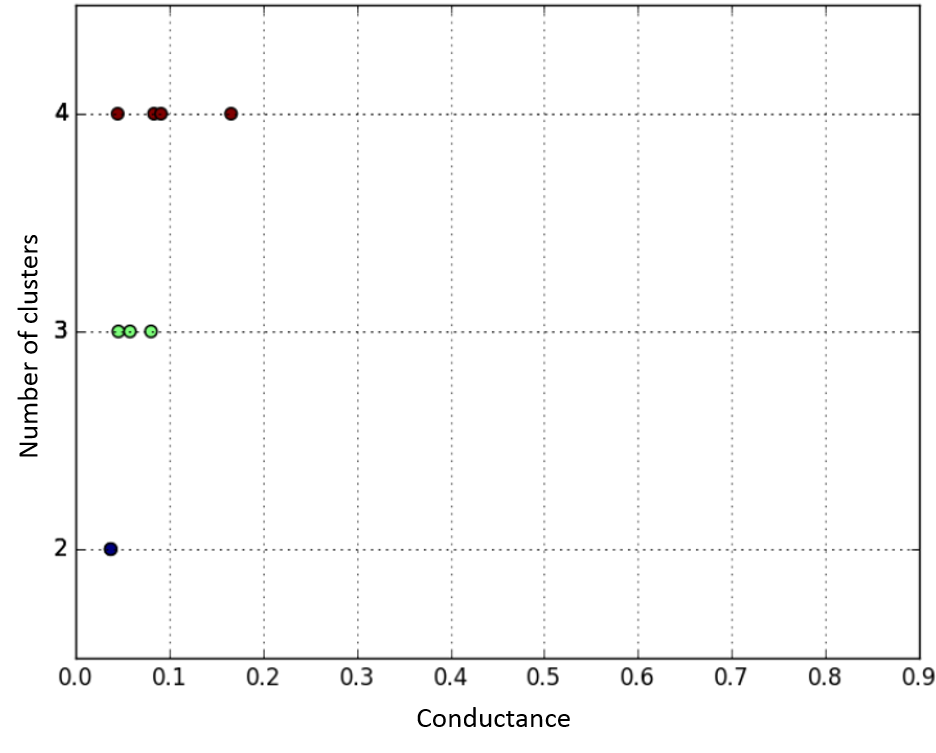}
    \caption{Conductance values for different number of clusters (value of $k$), for the topic {\em \q{2016 Brexit Referendum}}.}
    \label{fig:CondBrexit}
\end{figure}

Figure \ref{fig:CondBrexit} shows another example where three clusters are selected as representing different viewpoints. The plot corresponds to the topic {\em 2016 Brexit referendum} (more about the dataset in Section \ref{sec:evaluation}).
We notice that using $k=4$ for graph partitioning, three clusters have low conductance values (below 0.1). Since the fourth cluster has a much larger value, we can consider it as noisy and ignore it. As we will see in Section \ref{sec:evaluation}, one cluster corresponds to supporters of Brexit, one to supporters of Bremain, and the third to the neutral viewpoint where users generally discuss the consequences of Brexit and its relation to other topics. 

A simple algorithm for deciding both the number of clusters for partitioning the graph (value of $k$) and the number of clusters that hold different viewpoints is the following:  
We start by applying  the proposed MLGP clustering method for $k=2$ and inspecting the conductance value of the two clusters (which should be the same). If it is large (above a threshold $\delta$, e.g., $\delta = 0.10$), it means that the topic is probably not controversial or has not induced much discussion on the underlying social network, or the gathered tweets/retweets used for building the graph are not a good representation of the topic. 
On the contrary, if the value is below the threshold, then they are both selected as candidate clusters representing different viewpoints.
Then we partition the graph using $k=3$. If the conductance value of all three clusters is very low, this means that all clusters should be selected since they all hold a different viewpoint. If only two of the clusters have low conductance value (below the threshold), then we should better select $k=3$ and ignore the third \q{noisy} cluster. 
We continue in a similar way for larger values of $k$ until we get the maximum number of clusters below the threshold for the larger value of $k$.

\section{Viewpoint Understanding}
\label{sec:understanding}

To understand a viewpoint, we propose the iterative use of a simple rank difference method introduced in \cite{kit2008measuring}. We call the proposed method {\em Iterative Rank Difference}, for short IRD.

First, we try to answer the question: {\em what is a viewpoint about?}
Given a group of users $G_i \in G$ representing a specific viewpoint, we analyze the texts $P_i$ posted by the users of $G_i$ and derive a list of top-$n$ descriptive terms $W_i = \{w_{1}, \dots w_{n}\}$ that characterize the viewpoint. 
Then, to obtain a deeper understanding of a specific viewpoint, we try to answer the question: {\em what is a term about?} This allows understanding why a term appears in the list of a viewpoint's top descriptive terms. For example, as regards the {\em US 2016 presidential election}, we may want to understand why the hashtag {\em \#trumpleaks} appears in the top descriptive terms of the {\em \q{against Donald Trump}} viewpoint. Given a term $w \in W_i$, we answer this question by computing a set of other descriptive terms $W'_{i} = \{w'_{1}, \dots, w'_{n}\}$ that characterize $t$ in the context of $G_i$. 

\subsection{Rank Difference}
The rank difference algorithm tries to distinguish the important terms of a specific corpus and filter out terms that do not characterize the corpus \cite{kit2008measuring}. 
The algorithm operates over two ranked lists of terms: one that has been extracted from the \q{subject corpus} (corpus of interest related to a subject for which we want to find the descriptive terms), and one extracted from a \q{contrasting corpus} (a general corpus or corpus representing a different subject). Both lists are ranked according to the same criterion, e.g., the frequency of the terms in the corresponding corpus. Rank Difference identifies those terms with the largest difference in ranking between both corpora, in other words those terms that are most specific for the subject corpus.
The algorithm also considers the size of the lists in order to make the term ranks in the two corpora fairly comparable.

Formally, given the two ranked lists of terms $W_s$ and $W_c$ (one for the {\em subject} corpus and one for the {\em contrasting} corpus), the score of a term $w$ is computed as the difference in rank between the two corpora: 
\begin{equation}
    score(w) = \frac{rank(w, W_c)}{|W_c|} - \frac{rank(w, W_s)}{|W_s|} 
\end{equation}
where $rank(w, W_x) \in [-1,1]$ is the rank (position) of $w$ in the corresponding list. A term receives a high score if its position in the subject corpus is high and its position in the contrasting corpus is low (notice that the highest position in a list has rank 1). The higher the score of a term, the more important this term is for the subject corpus. By contrast, the lower the score of a term, the more important this term is for the contrasting corpus. 
If a term is in the same rank in the two lists and the lists have the same size, then its score is zero meaning that this term is not important, neither for the subject corpus nor for the contrasting corpus. 

Although normalization by the size of the lists makes the term ranks in the two corpora comparable, considering very large lists can affect the performance of this algorithm, e.g., by overemphasizing the importance of some terms with a low ranking criterion (e.g., frequency). For instance, if the top-100 terms of a list have a high frequency (e.g., $>1,000$) but there is a long tail of noisy terms with very low frequency (e.g., 10,000 terms with frequency $<10$), we should consider to cut the list in position 100.

\subsection{First iteration: {\em What is a viewpoint about?}}

To answer this question, we analyze the texts posted by all users in $U$ and create two lists of terms:
\begin{itemize}
    \item $W_{s}$ ({\em subject} list): Top-$n$ terms that occur in texts posted by the group of users $G_i$ representing the query viewpoint.
    \item $W_{c}$ ({\em contrasting} list): Top-$n$ terms that occur in texts posted by all other groups of users representing the other viewpoints.
\end{itemize}
In both lists, the term frequency (number of occurrences in the texts) is used as the ranking criterion. Then, we run the rank difference algorithm in these two lists and get the list of descriptive terms $W_i$ that characterize $G_i$.

\subsection{Next iterations: {\em What is a term about?}}
In the next iterations, we want to understand why one or more terms appear in the top-$n$ descriptive terms of a specific viewpoint. 
Given a term $w \in W_i$ that characterizes a viewpoint $G_i$, we run the rank difference algorithm on the following two lists: 

\begin{itemize}
    \item $W_s$ ({\em subject} list): Top-$n$ terms in texts mentioning $w$, posted by all users in $G_i$.
    \item $W_c$ ({\em contrasting} list): Top-$n$ terms in texts \underline{not} mentioning $w$, posted by all users in $G_i$.
\end{itemize}
The term frequency is used again as the ranking criterion in both lists.

\section{Evaluation}
\label{sec:evaluation}

\subsection{Effectiveness of Viewpoint Discovery}
We evaluated the effectiveness of the proposed viewpoint discovery method using the two Twitter datasets introduced in \cite{brigadir2015analyzing}\footnote{\label{foot:dataset}\url{http://dx.doi.org/10.6084/m9.figshare.1430449}}. The first dataset ({\em Indyref}) contains tweets about the 2014 Scottish Independence Referendum, where there are two main viewpoints: {\em Yes} (meaning support to Scottish independence), and {\em No} (meaning opposition to Scottish independence). The second dataset ({\em Midterms}) contains tweets about the 2014 U.S. Midterm Election. Here again, there are two main viewpoints: {\em Democratic party supporters}  and {\em Republican party supporters}. Details about the creation of the datasets and the ground truth can be found in  \cite{brigadir2015analyzing}.
Notice here that the ground truth for both datasets contains only users that belong to one of the two viewpoints, which however is not realistic. In a real setting, one has to collect a set of tweets and retweets related to a topic using some search terms, hashtags and/or user accounts, and then create a retweet graph which contains a very diverse and noisy set of users, many of who may not belong to a specific viewpoint. 
This makes the viewpoint discovery task much harder. However, since we are not aware of any other ground truth dataset that can be used for our problem (discovery of user-level viewpoints in a social network), we evaluate our approach using the aforementioned two datasets.\footnote{The SemEval 2016 Task 6 (Detecting Stance in Tweets) \cite{mohammad2016semeval} focuses on tweet-level stance detection.} 

We compare the proposed multi-level graph partitioning method (MLGP) with the best (for each dataset) unsupervised topic model proposed in \cite{Thonet2017Users}, which outperforms previous works on the same problem using the same datasets.
In order to have comparable results, we also discarded all the tweets with no interactions (replies or retweets) from the users in the dataset. 
We used {\em Purity} and {\em Normalized Mutual Information (NMI)} as the evaluation metrics. Purity measures the proportion of users who are assigned to the correct ground truth class, while NMI is based on mutual information and entropy \cite{danon2005comparing}. 

\begin{figure*}[h]
\centering
\subfloat[Indyref]{
\includegraphics[scale=.44]{figures/indyrefCond.png}
\label{fig:indyrefCond}
}
\subfloat[Midterms]{
\includegraphics[scale=.42]{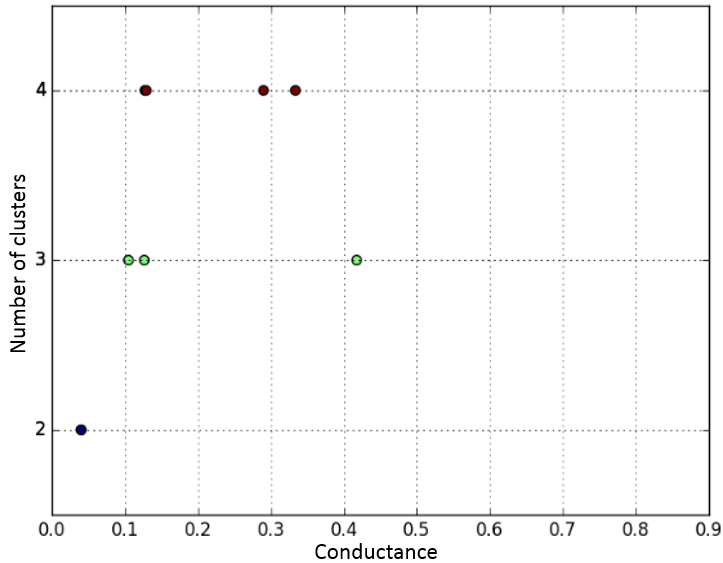}
\label{fig:midtermsCond}
}
\caption{Conductance values for different number of clusters (value of $k$), for the topics {\em Indyref} (a) and {\em Midterm} (b).}
\label{fig:condvalues}
\end{figure*}

Table \ref{tbl:effectivess} shows the results. The proposed MLGP method outperforms the best setting of the state-of-the-art topic model (SNVDM-GPU) in terms of both purity and NMI on both datasets. 
Figure \ref{fig:condvalues} shows the conductance plots of the two datasets. We notice that, for $k=2$ the produced clusters have very low conductance values of around 0.04 in both datasets, while for $k=3$ the values are highly increased, especially on the \textit{Indyref} dataset.

\begin{table}[]
\centering
\caption{Effectiveness of viewpoint discovery.}
\label{tbl:effectivess}
\begin{tabular}{@{}lllll@{}}
\toprule
\multicolumn{1}{c}{\multirow{2}{*}{Method}} & 
\multicolumn{2}{c}{Indyref} & 
\multicolumn{2}{c}{Midterms} \\
\multicolumn{1}{c}{} & \multicolumn{1}{c}{Purity} & \multicolumn{1}{c}{NMI} & \multicolumn{1}{c}{Purity} & \multicolumn{1}{c}{NMI} \\
\midrule
SNVDM-GPU &      0.969  &         0.800  &         0.964  &         0.778  \\
MLGP      & {\bf 0.988} & \textbf{0.908} & \textbf{0.983} & \textbf{0.876} \\
\bottomrule
\end{tabular}
\end{table}

\subsection{Qualitative Analysis of Viewpoint Understanding}
We used the proposed {\em Iterative Rank Difference (IRD)} method to try to understand the viewpoints of the following controversial topics: 
\begin{itemize}
    \item {\bf Indyref}: 2014 Scottish Independence Referendum
    \item {\bf USElection}: 2016 US Presidential Election
    \item {\bf Brexit}: 2016 Brexit referendum
\end{itemize}
For {\em Indyref}, we used the existing collection of tweets provided in \cite{kit2008measuring} (Footnote \ref{foot:dataset}), and we partitioned the graph using $k=2$.
For {\em USElection} and {\em Brexit}, we crawled topic-related tweets and retweets using the Twitter {\em Advanced Search} service and the retweet API. Tables \ref{tbl:uselectionQueries} and \ref{tbl:brexitQueries} show the hashtags and user accounts used for collecting the tweets related to these two topics. 
For the {\em USElection}, we used $k=3$ for graph partitioning and selected the two clusters with the lowest conductance values as those representing different viewpoints,
while for {\em Brexit} we used $k=4$ and selected three clusters (this selection is based on the method described in Section \ref{sec:discovery}).
For creating the {\em subject} ($W_s$) and {\em contrasting} ($W_c$) lists of top-$n$ terms required by IRD, we preprocessed the texts of the tweets, removed user mentions, stopwords, punctuations, 1- and 2-char words, and URLs, and lemmatized the remaining terms. Moreover, we used $n=200$ to cut the subject and contrasting lists in all cases except the 2nd iteration of the Indyref topic where we used $n=50$ (for Indyref the data is less and the term frequencies smaller compared to the other two topics).

\begin{table}[h]
\centering
\caption{Hashtags and user accounts used for collecting the tweets related to {\em USElection}.}
\vspace{-2mm}
\label{tbl:uselectionQueries}
\begin{tabular}{p{8cm}}
\toprule
\#USElection, \#USElection2016, \#trump, \#MakeAmericaGreatAgain,
@DaysOfTrump, \#TrumpLeaks, \#obamacare, \#obama, \#realDonaldTrump,
\#trump2016, \#nevertrump, \#Clinton, \#Hillary, \#hillaryClinton, \#USElection,
\#BARACKOBAMA, \#gop, \#Republican, \#ObamaFarewell, \#OurFirstStand,
\#MAGA, \#fakenews, @realdonaldtrump, @hfa @GOP,
@BarackObama, @HillaryClinton, @TheDemocrates  \\   
\bottomrule
\end{tabular}
\vspace{-3mm}
\end{table}

\begin{table}[h]
\centering
\caption{Hashtags used for collecting the tweets related to {\em Brexit}.}
\vspace{-2mm}
\label{tbl:brexitQueries}
\begin{tabular}{p{8cm}}
\toprule
\#brexit, \#notoeu, \#betteroffout, \#britainout, \#leaveeu, \#beleave,
\#loveeuropeleaveeu, \#yestoeu, \#betteroffin, \#votein, \#bremain, \#strongerin,
\#leadnotleave, \#voteremain, \#stopbrexit \\   
\bottomrule
\end{tabular}
\end{table}

\subsubsection{What is a viewpoint about}

In the first iteration, our objective is to understand what each viewpoint represents.
For each topic, Tables \ref{tbl:indyrefTop10}-\ref{tbl:BrexitTop10} show the top descriptive terms of the corresponding viewpoints, as derived by the proposed method. 

Regarding {\em Indyref}, we notice that the first viewpoint (Viewpoint 1) clearly represents the supporters of \q{Yes}, including terms like the hashtags {\em \#voteyes}, {\em \#yes}, and the words {\em independent} and {\em national}. 
The second viewpoint (Viewpoint 2) clearly represents the supporters of \q{No} containing many related terms like {\em \#nothanks}, {\em \#bettertogether}, {\em \#labourno}, and {\em \#voteno}. Moreover, {\em currency} seems to be an important topic for this viewpoint.

In the {\em USElection} topic, the first viewpoint (Viewpoint 1) represents the supporters of Hillary Clinton (or the opponents of Donald Trump), while the second (Viewpoint 2) the supporters of Donald Trump. We notice that the supporters of Hillary Clinton used hashtags like {\em \#nevertrump}, \textit{\#theresistance}, \textit{\#imwithher}, and \textit{\#trumpleaks} to oppose to Donald Trump, while Trump's supporters use hashtags like \textit{\#trump2016}, \textit{\#trumptrain}, \textit{\#draintheswamp}, \textit{\#makeamericagreatagain}, and \textit{\#fakenews}. 

For {\em Brexit}, we see that the first viewpoint (Viewpoint 1) represents Brexit supporters (\textit{\#go, \#leave, free, control, \#leaveeu}), the second (Viewpoint 2) characterizes Bremain supporters (\textit{\#votein, \#stopbrexit}) and their worry about the National Health Service (\textit{\#nhs, cost}), while the third (Viewpoint 3) reflects a more neutral viewpoint about the general consequences of Brexit (\textit{\#stocks, \#healthinnovations, \#banking, \#pharma, global}) and its relation to other topics (\textit{\#maga, \#cdnpoli, \#trumptrain}), probably discussed by users of non-English languages (\textit{\#ue, unido, reino, londres}).  
We also notice that Viewpoint 2 (Bremain supporters) contains the terms {\em \#theresamay} and {\em johnson} which characterize Brexit (Theresa May and Boris Johnson belong to the Conservative Party which supported Brexit). This is not surprising since users supporting a specific viewpoint may use terms and hashtags that characterize another viewpoint for criticizing it. 

\begin{table}[]
\centering
\caption{Top descriptive terms of viewpoints about the {\em 2014 Scottish Independence Referendum}.}
\label{tbl:indyrefTop10}
\begin{tabular}{p{3.5cm}p{4.1cm}}
    \toprule
    Viewpoint 1 & Viewpoint 2   \\
    \midrule
    \#voteyes, \#yes, westminster, meeting, independent, \#scotland, murphy, event, national, folk
    &  
    \#nothanks, \#bettertogether, currency, \#labourno, \#scotdecides, speech, alex, \#voteno, part, seperation \\
  \bottomrule
\end{tabular}
\end{table}

\begin{table}[]
\centering
\caption{Top descriptive terms of viewpoints about the {\em 2016 US Presidential Election}.}
\label{tbl:USElectionTop10}
\begin{tabular}{p{3.7cm}p{4.1cm}}
    \toprule
    Viewpoint 1 & Viewpoint 2   \\
    \midrule
    \#nevertrump, \#theresistance, \#obamafarewell, \#imwithher, \#resist, \#trumps, gop, tweet, \#gop, \#trumpleaks, tax, \#notmypresident, putin
    &  
    \#tcot, \#trump2016, \#pjnet, \#trumptrain, \#draintheswamp, video, \#makeamericagreatagain, \#fakenews, breaking, god, \#realdonaldtrump, usa, fbi \\
  \bottomrule
\end{tabular}
\end{table}

\begin{table}[]
\centering
\caption{Top descriptive terms of viewpoints about the {\em 2016 Brexit referendum}.}
\label{tbl:BrexitTop10}
\begin{tabular}{p{2.0cm}p{1.9cm}p{3.3cm}}
    \toprule
    Viewpoint 1 & Viewpoint 2 & Viewpoint 3  \\
    \midrule
    democracy, \#nexit, try, \#go, \#referendumm, \#leave, \#britain, free, control, \#leaveeu
    &   
    \#tory, tory, johnson, \#theresamay, boris, \#votein, \#stopbrexit, \#nhs, cost, nhs
    & 
    \#ue, unido, reino, \#maga, \#stocks, \#cdnpoli, europa, londres, \#americafirst,
    royaumeuni, \#trumptrain, \#healthinnovations, \#banking, \#pharma \\ 
  \bottomrule
\end{tabular}
\end{table}

\subsubsection{What is a term about?}
In the next iterations, our objective is to find more information about specific terms that characterize a viewpoint. This can help to obtain a deeper understanding of the viewpoint and its related context.

In the {\em Indyref} topic, Table \ref{tbl:termAbout} shows the top-5 descriptive terms of {\em murphy} and {\em \#voteyes} (Viewpoint 1). Regarding {\em murphy}, we notice that the list provides context information about Jim Murphy: Jim Murphy was the {\em leader} of the Scottish {\em Labour} Party during the referendum, suggested that Labour could offer more {\em tax} and welfare powers for Scotland, while he is also believed to have supported the Iraq {\em war}.
As regards {\em \#voteyes}, we notice that the list does not provide much information. The reason is that there is also the hashtag {\em \#yes} which semantically is the same with {\em \#voteyes}. Both hashtags were interchangeably used by Twitter users to indicate their support to Scottish Independence, meaning that there are almost no tweets containing both hashtags. Thus, by creating the {\em subject} list $W_s$ without considering the similar hashtag {\em \#yes}, we indirectly consider this term as belonging to the {\em contrasting} corpus. 
Table \ref{tbl:termAbout2} shows the top-5 descriptive terms when considering both hashtags in the {\em subject} corpus. Now we notice that the list contains terms that characterize the support to \q{Yes}: an {\em independent Scotland} will give {\em power} and {\em future} to the {\em country}.
Studying methods for the identification of identical terms is beyond the scope of this paper but an interesting direction for future research. 

\begin{table}[]
\centering
\caption{Top-5 descriptive terms of {\em murphy} and {\em \#voteyes} (Viewpoint 1 of \textit{Indyref}).}
\vspace{-1mm}
\label{tbl:termAbout}
\begin{tabular}{p{3.0cm}p{4.5cm}}
    \toprule
    \centering\arraybackslash murphy &  \centering\arraybackslash \#voteyes   \\
    \midrule
    leader, war, tax, labour, really      
    &  \#scotland, future, \#sexysocialism, \#yesscot, \#midlothiansaysyes  \\ 
  \bottomrule
\end{tabular}
\end{table}

\begin{table}[]
\centering
\caption{Top-5 descriptive terms of both {\em \#voteyes} and {\em \#yes} (Viewpoint 1 of {\em Indyref}).}
\vspace{-1mm}
\label{tbl:termAbout2}
\begin{tabular}{p{7.0cm}}
    \toprule
    \centering\arraybackslash \{\#voteyes, \#yes\}   \\
    \midrule
    scotland, independent, power, future, country      \\ 
  \bottomrule
\end{tabular}
\end{table}

Regarding the {\em USElection} topic, Table \ref{tbl:termAboutUSElection} shows 
the top-5 descriptive terms of {\em \#trumpleaks} (Viewpoint 1) and {\em fbi} (Viewpoint 2). 
For {\em \#trumpleaks} we see that this term is related to a report authored by Scott Dworkin ({\em \#dworkinreport}) which shows that Donald Trump has incorporated many registered businesses in Russia ({\em \#trumprussia}), while the opponents  of Donald Trump ask Congress to investigate whether sufficient grounds exist for his impeachment ({\em \#impeachtrump}). 
As regards {\em fbi}, we notice that this term is related to the {\em reopening} of an {\em investigation/case} as well as to {\em James Comey} who served as the seventh {\em director} of FBI during the US election. Indeed, James Comey reopened an FBI investigation into Hillary Clinton's use of private email server while she was secretary of State. 

\begin{table}[]
\centering
\caption{Top-5 descriptive terms of {\em \#trumpleaks} (Viewpoint 1) and {\em fbi} (Viewpoint 2) of \textit{USElection}.}
\vspace{-1mm}
\label{tbl:termAboutUSElection}
\begin{tabular}{p{5.0cm}p{2.8cm}}
    \toprule
    \centering\arraybackslash \#trumpleaks &  \centering\arraybackslash fbi   \\
    \midrule
    \#trumprussia, \#impeachtrump, \#dworkinreport, \#amjoy, \#muslimban   
    &   
    investigation, comey, reopen, director, case  \\ 
  \bottomrule
\end{tabular}
\end{table}

Finally, in the {\em Brexit} topic, Table \ref{tbl:termAboutBrexit} shows the top-5 descriptive terms of {\em \#nexit} (Viewpoint 1), {\em control} (Viewpoint 1), {\em \{boris, johnson\}} (Viewpoint 2), and {\em banking} (Viewpoint 3).
We notice that the top descriptive terms of {\em \#nexit} (which is the hashtag used for the hypothetical Dutch withdrawal from the European Union) is related to similar hashtags used for other countries (France, Italy, Denmark, Sweden) as well as {\em \#euistheproblem}, meaning that the supporters of Brexit contravene European Union. 
As regards {\em control}, it is clear that the supporters of Brexit believe that UK must take back the total control of the country as well as of its borders.
Regarding {\em \{boris, johnson\}} (Viewpoint 2) we get general information about Boris Johnson who is member of the Conservative Party in the United Kingdom ({\em conservative}). 
Boris Johnson has been Secretary of State for Foreign and Commonwealth Affairs since 2016 ({\em \#foreignsecratery}) and the Member of Parliament (MP) for Uxbridge and South Ruislip since 2015 ({\em \#mp}). We also see the term {\em biscuit} which is related to his visit to a biscuit factory during his \q{Vote Leave} campaign.  
Finally, the top descriptive terms about {\em banking} (Viewpoint 3) relate this viewpoint to the effect of Brexit to UK stocks ({\em \#stocks, decrease, index}) and the pharma industry ({\em \#healthinnovations, \#pharma}).

\begin{table}[]
\centering
\caption{Top-5 descriptive terms of {\em \#nexit} (Viewpoint 1), {\em control} (Viewpoint 1),  {\em \{boris, johnson\}} (Viewpoint 2), and {\em banking} (Viewpoint 3) of \textit{Brexit}.}
\label{tbl:termAboutBrexit}
\begin{tabular}{p{1.6cm}p{1.6cm}p{2.0cm}p{1.8cm}}
    \toprule
    \centering\arraybackslash \#nexit & control &  \centering\arraybackslash \{boris, johnson\} & banking \\
    \midrule
    \#frexit, \#italexit, \#dexit, \#swexit, \#euistheproblem
    &
    \#takecontrol, border, uncontrolled, \#betteroffout, \#takebackcontrol
    &   
    conservative, biscuit, \#foreignsecratery, \#mp, \#borisjohnson
    &  
    \#stocks, \#healthinnovations, \#pharma, decrease, index
      \\ 
  \bottomrule
\end{tabular}
\end{table}

\section{Conclusion}
\label{sec:conclusion}
We have proposed to combine a popular graph partitioning algorithm with a clustering quality metric for the problem of viewpoint discovery in social networks. A distinctive characteristic of the proposed method is that it does not require the number of viewpoints to be given as input. This makes our approach applicable also for cases with an unknown number of viewpoints, while it can also detect noisy groups of users that do not represent clear viewpoints. Evaluation results on publicly available ground truth datasets showed that our approach outperforms state-of-the-art topic models on the same problem. 

To understand what a viewpoint is about, we proposed the iterative use of a simple rank difference algorithm. The introduced {\em Iterative Rank Difference} (IRD) method can automatically identify descriptive terms that characterize a viewpoint, allowing also to understand how a specific term is related to a viewpoint. 
Case studies on three different controversial topics showed that IRD can provide comprehensive representations of viewpoints and terms.

Regarding future research, we plan to study approaches for the \textit{timeline summarisation} of topics and viewpoints that will allow understanding how a controversial topic evolves over time and with respect to the involved entities, events and subtopics. 
Another interesting direction is the {\em semantic representation} of topics and viewpoints which will enable the construction of queryable knowledge graphs about controversial topics.

\begin{acks}
The work was partially funded by the European Commission for the ERC Advanced Grant ALEXANDRIA (No. 339233).
\end{acks}

\bibliographystyle{ACM-Reference-Format}
\balance
\bibliography{WebSci18__BIB}
\end{document}